\documentclass[prl,final,twocolumn,superscriptaddress]{revtex4}
\usepackage{amsmath}
\usepackage{color}
\usepackage{epstopdf}
\usepackage{natbib}
\usepackage[normalem]{ulem}
\usepackage[pdftex]{graphicx}
\usepackage{times}
\usepackage{txfonts}
\usepackage{textcomp}
\usepackage[colorlinks, bookmarks=false, citecolor=blue, linkcolor=red, urlcolor=blue]{hyperref}

\def\be {\begin{equation}}
\def\ee {\end{equation}}

\def\horparallel{ \lower.5ex\hbox{ \includegraphics[width=2ex]{fig-hor.pdf}}\,\, }
\def\vertparallel{ \lower.5ex\hbox{ \includegraphics[width=2ex]{fig-vert.pdf}}\,\, }
\def\gsim{\mathrel{\rlap{\lower4pt\hbox{\hskip1pt$\sim$}}\raise1pt\hbox{$>$}}}
\def\lsim{\mathrel{\rlap{\lower4pt\hbox{\hskip1pt$\sim$}}\raise1pt\hbox{$<$}}}

\begin{document}
\title{Quantum stabilization of classically unstable plateau structures}
\author{Tommaso Coletta}
\affiliation{Institute of Theoretical Physics, Ecole Polytechnique F\'ed\'erale de Lausanne (EPFL),
CH-1015 Lausanne, Switzerland}
\author{M. E. Zhitomirsky}
\affiliation{Service de Physique Statistique, Magn\'etisme et Supraconductivit\'e, UMR-E9001 CEA-INAC/UJF,
17 rue des Martyrs, F-38054 Grenoble, France}
\author{Fr\'ed\'eric Mila}
\affiliation{Institute of Theoretical Physics, Ecole Polytechnique F\'ed\'erale de Lausanne (EPFL),
CH-1015 Lausanne, Switzerland}
\date{\today}

\begin{abstract}
Motivated by the intriguing report, in some frustrated quantum antiferromagnets, of magnetization
plateaus whose simple collinear structure is {\it not} stabilized by an external magnetic field in
the classical limit, we develop a semiclassical method to estimate the zero-point energy of collinear
configurations even when they do not correspond to a local minimum of the classical energy. For
the spin-$1/2$ frustrated square-lattice antiferromagnet, this approach leads to the stabilization of
a large 1/2 plateau with ``up-up-up-down'' structure for $J_2/J_1>1/2$, in agreement with exact
diagonalization results, while for the spin-$1/2$ anisotropic triangular antiferromagnet, it predicts that
the 1/3 plateau with ``up-up-down'' structure is stable far from the isotropic point, in agreement with
the properties of Cs$_2$CuBr$_4$.
\end{abstract}

\maketitle

{\it Introduction.---}%
Frustration is responsible for the emergence of several remarkable properties in quantum magnets,
ranging from rather exotic types of order such as quadrupolar or nematic order to resonating
valence bond or algebraic spin liquids \cite{Introduction_FM}. In the presence of an external field,
frustration is also known to be at the origin of several types of accidents in the magnetization curve,
including kinks, jumps and plateaus. Of all these remarkable features, magnetization plateaus at
rational value of the magnetization are probably the best documented ones experimentally, and their
theory is likewise quite advanced. Following the terminology of Hida and Affleck \cite{hida},
two kinds of plateaus have been identified \cite{takigawa_mila}:
`classical' plateaus \cite{Kawamura84,Chubukov-Golosov,Zhitomirsky-Honecker-Petrenko}, whose structure
has a simple classical analog with spins up or down along the external field, and `quantum' plateaus
\cite{totsuka,mila,miyahara,Dorier08,abendschein}, which have no classical analog and correspond to
a Wigner crystal of
triplets in a sea of singlets. In the case of quantum plateaus, the mechanism is clear: frustration
reduces the kinetic energy of triplets, resulting in a crystallization at commensurate densities.
The main open problem is to be predictive for high commensurability plateaus since it requires
a precise knowledge of the long-range part of the triplet-triplet interaction.

By contrast, and somehow surprisingly, the theory of classical plateaus is not
complete yet. The paradigmatic example of a classical plateau is the 1/3 magnetization plateau
of the Heisenberg antiferromagnet on a triangular lattice,
studied by Chubukov and Golosov \cite{Chubukov-Golosov} in the context of a $1/S$ expansion.
In this system the three sublattice up-up-down (\textit{uud}) structure
appears classically at $H=H_{\textrm{sat}}/3$, and quantum fluctuations stabilize this \textit{uud} state
in a finite field range around $H_{\textrm{sat}}/3$, leading to the 1/3 plateau. The basic qualitative idea
in the spirit of the order by disorder
is that collinear configurations often have a softer spectrum and, hence, a smaller zero-point energy
\cite{Shender,Henley}.
The prediction of the 1/3 plateau has been confirmed by exact diagonalization of finite clusters for $S=1/2$ and
1 \cite{Honecker04}, and the theory of Chubukov and Golosov can be extended to all cases where a collinear
state is classically stabilized for a certain field.

There are cases, however, where a classical plateau has been suggested to exist although
the collinear structure stabilized
for quantum spins is {\it not} the ground state for classical spins, the classical ground state
in the appropriate field range being in general a non-coplanar structure. This is
for instance the case of the spin-$1/2$ $J_1$--$J_2$ model on the square lattice, for which
exact diagonalizations have revealed the presence of a four sublattice up-up-up-down (\textit{uuud})
1/2 plateau in a parameter range where the classical ground state has a canted stripe structure
(see below). Another example is the 1/3 plateau of the Heisenberg model on the anisotropic triangular
lattice, a model relevant to the compound Cs$_2$CuBr$_4$. To develop a general theory of classical
plateaus in that situation remains the main open issue in the field.

The goal of this Letter is to develop such a theory. For that purpose, we start with a general
Heisenberg model in an external field defined by the spin Hamiltonian
\begin{equation}\label{eq:Generic Heisenberg model}
 \mathcal{H}= \sum_{\langle i,j\rangle} J_{ij}\, {\bf S}_i \cdot {\bf S}_j -H\sum_i S_i^z,
\end{equation}
and show how to estimate the zero-point energy of collinear states {\it even if they do not
minimize the classical energy}. More precisely, we derive an upper bound of this energy to order
$1/S$. If the energy of a collinear state estimated in this way is lower than that of the classical
ground state (including the zero-point energy), then the collinear state
must be the ground state since the energy used for the comparison is an {\it upper bound}.
Correspondingly, the quantum antiferromagnet exhibits a magnetization plateau in
a certain field range, which may generally exceed our conservative theoretical estimate based
on the upper energy-bound. We apply this approach to the $J_1$--$J_2$ model on a square lattice and
to the Heisenberg antiferomagnet on the anisotropic triangular lattice,
with results in remarkable agreement with existing numerical data for $S=1/2$.

{\it General formalism.---}%
We propose a simple method to test for the stability of plateau structures in the framework of the linear
spin-wave theory. First of all, we note that, since plateau structures are collinear, the fluctuation
Hamiltonian around such structures in terms of Holstein-Primakoff bosons does not contain linear bosonic terms,
and that the first relevant terms in its $1/S$ expansion are quadratic. Since plateau structures are classical
minima of the energy only at specific values of couplings and magnetic field (if any) the corresponding
harmonic fluctuation Hamiltonians are positive definite only at these specific points.
Away from such points the correction to the classical energy cannot be straightforwardly computed. The fact that
the spectrum is not well defined stems from the harmonic approximation. If the
plateau state is to become the true quantum ground state,  higher order terms in the spin-wave expansion
must produce an excitation spectrum with positive frequencies.
This approach requires to calculate nonlinear quantum corrections to the spectrum as
done in Ref.~\cite{Chubukov-Golosov} and in more recent studies \cite{Alicea-Chubukov-Starykh,Takano11}.
This is a rather cumbersome procedure, and it would be useful
to have a simpler approach to determine if there is a plateau, and to estimate its
width to the lowest-order in $1/S$. Besides, the calculation of the excitation spectrum
for the plateau region allows to identify only second-order transitions, while in
many experimental and model examples the transitions at the plateau boundaries are
of the first-order and, therefore, require a full energetic comparison.

The proposed method to obtain a well-defined spectrum around a state which is not
a classical ground state consists in adding a staggered-field term
\begin{equation}\label{eq:V}
 \hat{V}=\delta\sum_i\left(S-S_i^{z_i}\right)
\end{equation}
to the harmonic Hamiltonian. A similar approach has been introduced in a different
context in \cite{Coletta-Korshunov-Mila,Wenzel-Coletta-Korshunov-Mila}.
On each lattice site $i$, the staggered field $\delta>0$ is oriented in the direction
$\hat{\bf z}_i$ of the corresponding classical spin. The extra term \eqref{eq:V}
amounts to a shift of the chemical potential of the Holstein-Primakoff bosons
$\hat{V}=\delta\sum_i a_i^\dagger a_i$ and yields a positive contribution to
the spin-wave Hamiltonian. The magnitude of $\delta$ is adjusted to ensure that the harmonic
Hamiltonian is positive definite. The resulting spectrum obtained with the help of
the Bogolyubov transformation has real and positive frequencies. The advantages of this
variational approach are the following: the addition of $\hat{V}$ to the Hamiltonian
does not change the classical energy of the trial state
and allows to obtain physically meaningful dispersion relations.
Furthermore, since the expectation value of $\hat{V}$ is strictly positive,
the computed energy correction provides an {\it upper} bound for the energy of
the plateau state.
It should be noted that the suggested method is valid not only for collinear spin structures
but can be extended to all structures that are saddle points of the classical energy, see
\cite{Coletta-Korshunov-Mila,Wenzel-Coletta-Korshunov-Mila} and below.

{\it $J_1$--$J_2$ model.---}%
For the frustrated square-lattice antiferromagnet the exchange interaction constants are
$J_{ij}=J_1$ and $J_2$ for nearest and second-nearest neighbors, respectively. In zero field the classical
ground state of the model is a helix with the ordering wave vector given by the minimum of the Fourier
transform of the coupling interaction
$J_{\bf q}=4J_1\gamma_{\bf q}+4J_2\eta_{\bf q}$ with
$\gamma_{\bf q}=(\cos{q_x}+\cos{q_y})/2$ and $\eta_{\bf q}=\cos{q_x}\cos{q_y}$.
For $J_2/J_1 < 1/2$, the minimum corresponds to
${\bf q}_N=(\pi,\pi)$, i.e. to N\'eel order. In the opposite case
$J_2/J_1>1/2$, the order by disorder mechanism selects collinear striped structures with ordering wave vectors
${\bf q}_S=(\pi,0)$ or $(0,\pi)$ \cite{Henley}.
The point $J_2/J_1=1/2$ is highly degenerate since $J_{\bf q}$ is minimal along the lines $q_x=\pi$ and $q_y=\pi$.
In the presence of a magnetic field both the N\'eel and the stripe structure are canted with a uniform spin component
in the field direction. The canting angle $\theta$ measured with respect to the $z$ axis is given by $\cos\theta_N= H/8J_1S$
and $\cos\theta_S= H/(4J_1+8J_2)S$ for the two states.

The analysis of classical spin configurations in a magnetic field suggests the appearance of
a 1/2-magnetization plateau with a four-sublattice $uuud$ structure for the strongly frustrated
point $J_2/J_1=1/2$ \cite{Zhitomirsky-Honecker-Petrenko}. The conclusion has been supported by
exact diagonalizations of finite clusters, though numerically the plateau extends well into
the classically unstable region  $J_2/J_1>1/2$ with the largest width at $J_2/J_1\approx 0.6$.
Moreover, the linear spin-wave calculation for $J_2/J_1=1/2$ shows that for this ratio of coupling constants
the canted N\'eel state  wins over the collinear plateau state \cite{Jackeli-Zhitomirsky}
leaving an apparent problem with reconciling numerical and analytical results.

We now investigate the appearance of the 1/2-magnetization plateau for the
$J_1$--$J_2$ model using the variational harmonic theory outlined above. In the semiclassical approach
deviations from the classical configuration are expressed as Holstein Primakoff
bosons \cite{Holstein-Primakoff}. In the harmonic approximation the bosonic Hamiltonian
can be split into three contributions:
\begin{equation}
 \mathcal{H}=\mathcal{H}^{(0)}+\mathcal{H}^{(1)}+\mathcal{H}^{(2)},
\end{equation}
where $\mathcal{H}^{(0)}$ is the classical energy of the system. $\mathcal{H}^{(1)}$ and $\mathcal{H}^{(2)}$
respectively contain only terms which are linear and quadratic in boson operators.
The \textit{uuud} state, being a collinear state, is such that no linear terms appear when performing
the HP transformation, hence $\mathcal{H}^{(1)}=0$ for this state. In non collinear structures
$\mathcal{H}^{(1)}$ is proportional to the derivative of the classical energy with respect to spin orientations.
It vanishes for the canted N\'eel and canted stripe structures over the entire parameter range since both
are extrema of the classical energy. In fact the canted N\'eel (stripe) structure is a saddle point of
the classical energy for $H<H_\textrm{sat}$ and $J_2/J_1>1/2$ (respectively $J_2/J_1<1/2$).

The general structure of the quadratic bosonic Hamiltonian that describes harmonic fluctuations around these
various configurations is given by
\begin{equation}\label{eq:H sw}
 \mathcal{H}=  N E_\textrm{cl} + \frac{1}{2} \sum_{\bf k} \Bigl[
 \hat{\bf a}_{\bf k}^\dagger M_{\bf k}\hat{\bf a}_{\bf k}-\Delta_{\bf k} \Bigr]\ ,
\end{equation}
where $E_{\textrm{cl}}$ is the classical energy per site of the state around which fluctuations are considered.
For the canted N\'eel and the canted stripe states
$\hat{\bf a}_{\bf k}^\dagger = (a_{\bf k}^\dagger,a_{-\bf k})$ and $M_{\bf k}$ is the $2\times2$ matrix
\begin{equation}\label{eq:M}
M_{\bf k}({\bf q})=\left(\begin{array}{cc}
			      A_{\bf k}({\bf q}) & B_{\bf k}({\bf q}) \\
			      B_{\bf k}({\bf q}) & A_{\bf k}({\bf q})
			      \end{array}\right).
\end{equation}
Coefficients $A_{\bf k}({\bf q})$ and $B_{\bf k}({\bf q})$ for N\'eel and striped structures
are listed below:
\begin{eqnarray}
 A_{\bf k}({\bf q}_N) & = & 4J_1S(1 + \gamma_{\bf k} \cos^2\theta_N)-4J_2S(1-\eta_{\bf k})\ ,
 \nonumber \\
 B_{\bf k}({\bf q}_N) & = & -4J_1S \gamma_{\bf k} \sin^2\theta_N
\label{eq:AB neel}
\end{eqnarray}
and
\begin{eqnarray}
 A_{\bf k}({\bf q}_S\!)& = & 4J_2S(1\!+\eta_{\bf k}\!\cos^2\!\theta_S\!)+2J_1S (\cos{k_y}\!+\cos^2\!\theta_S\cos{k_x}\!) ,
 \nonumber \\
 B_{\bf k}({\bf q}_S\!) & = & -2S\sin^2\theta_S(J_1\cos k_x+2J_2\eta_{\bf k})
 \label{eq:AB stripe}
\end{eqnarray}
where we have used ${\bf q}_S=(\pi,0)$. The additional constants $\Delta_{\bf k}^N$ and
$\Delta_{\bf k}^S$ in Eq.~\eqref{eq:H sw} are given respectively by $A_{\bf k}({\bf q}_N)$ and
$A_{\bf k}({\bf q}_S)$.

\begin{figure}[t]
\centering
\includegraphics[width=0.8\columnwidth]{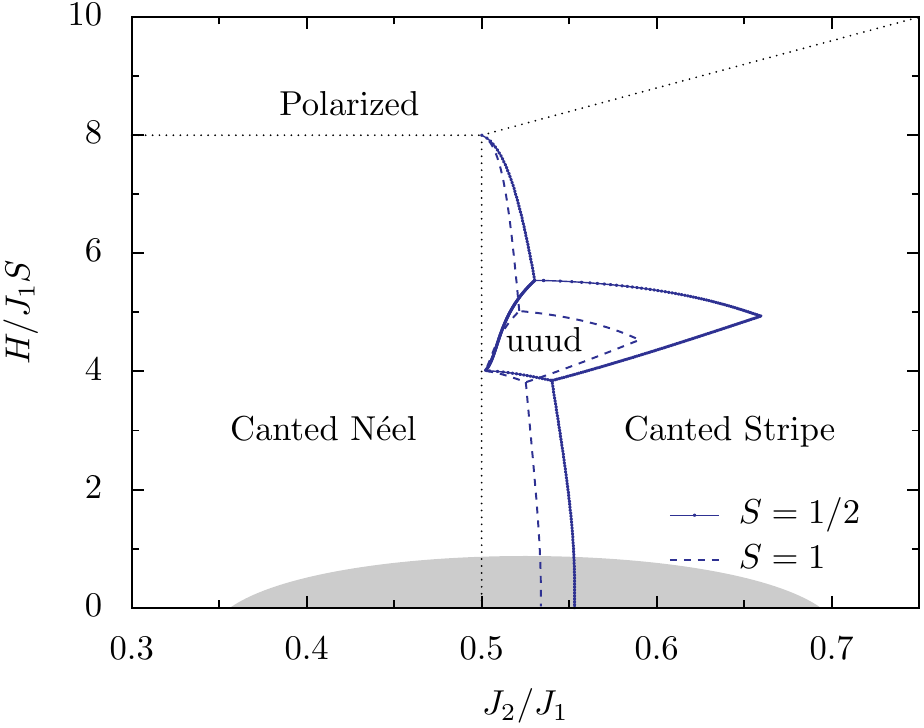}
\caption{Semiclassical phase diagram for the spin $1/2$ $J_1-J_2$ Heisenberg model in magnetic
field. The $uuud$ structure is stabilized by fluctuations over a wide parameter range. Dashed lines
correspond to the phase diagram for $S=1$ and dotted lines to the classical phase boundaries.
The shaded area represents schematically the gapped singlet phase for the spin-1/2 model.}
\label{fig:J1J2 phase diagram}
\end{figure}

The \textit{uuud} state has a four-site unit cell, and $\hat{\bf a}_{\bf k}^\dagger$ denotes
$(a_{1,{\bf k}}^\dagger,\ldots,a_{4,{\bf k}}^\dagger,a_{1,-{\bf k}},\ldots,a_{4,-{\bf k}})$ with
$M_{\bf k}$ being the $8\times8$ matrix obtained from (\ref{eq:M}) by substituting
\begin{equation}
\begin{array}{ll}
 A_{\bf k}=\left(
\begin{array}{cccc}
\bar{A}_{\bf k} & \bar{E}_{\bf k} & 0 & \bar{G}_{\bf k}^\star \\
\bar{E}_{\bf k}^\star & \bar{B}_{\bf k} & 0 & \bar{F}_{\bf k} \\
0 & 0 & \bar{C}_{\bf k} & 0 \\
\bar{G}_{\bf k} & \bar{F}_{\bf k}^\star & 0 & \bar{A}_{\bf k} \\
\end{array}
\right)\ ,&
 B_{\bf k}=\left(
 \begin{array}{cccc}
 0 & 0 & -\bar{F}_{\bf k} & 0 \\
 0 & 0 & \bar{H}_{\bf k}^\star & 0 \\
 -\bar{F}_{\bf k}^\star & \bar{H}_{\bf k} & 0 & -\bar{E}_{\bf k} \\
 0 & 0 & -\bar{E}_{\bf k}^\star & 0 \\
 \end{array}
\right)
\end{array}
\end{equation}
with coefficients
\begin{equation}
\begin{array}{l}
\begin{array}{ll}
\bar{A}_{\bf k}=-4J_2S+H\ , & \bar{B}_{\bf k}=-4(J_1-J_2)S+H\ , \\ [3mm]
\bar{C}_{\bf k}=4(J_1+J_2)S-H\ , & \bar{E}_{\bf k}=J_1S\tau_{k_y}\ , \\[3mm]
\end{array}\\[3mm]
\begin{array}{lll}
\bar{F}_{\bf k}=J_1S\tau_{k_x}\ ,& \bar{G}_{\bf k}=J_2S\tau_{-k_x}\tau_{-k_y}\ ,& \bar{H}_{\bf k}=-J_2S\tau_{-k_x}\tau_{k_y}\ ,
\end{array}
\end{array}
\end{equation}
where $\tau_k=(1+e^{-i2k})$.
The constant $\Delta_{\bf k}^\textrm{\textit{uuud}}$ is given by $2\bar{A}_{\bf k}+\bar{B}_{\bf k}+\bar{C}_{\bf k}$.
When the quadratic form of Eq.~(\ref{eq:H sw}) is positive definite, it can be diagonalized by the standard Bogolyubov
transformation allowing to compute the quantum corrections to the classical energy.
For parameters for which the matrix $M_{\bf k}$ is not positive definite we add to the Hamiltonian the term
$\hat{V}$ defined in Eq.~(\ref{eq:V}). From the expression of $\hat{V}$ in terms of HP bosons it is clear that
it leaves $\mathcal{H}^{(0)}$ and $\mathcal{H}^{(1)}$ unchanged while its effect on the bosonic Hamiltonian is
to increase $\Delta_{\bf k}$ and all diagonal elements of $M_{\bf k}$ by $\delta/2$. The field $\delta$ is adjusted
to the minimal value which is sufficient to make $M_{\bf k}$ positive definite over the entire Brillouin zone.

\begin{figure}[h!]
\centering
\includegraphics[width=1.0\columnwidth]{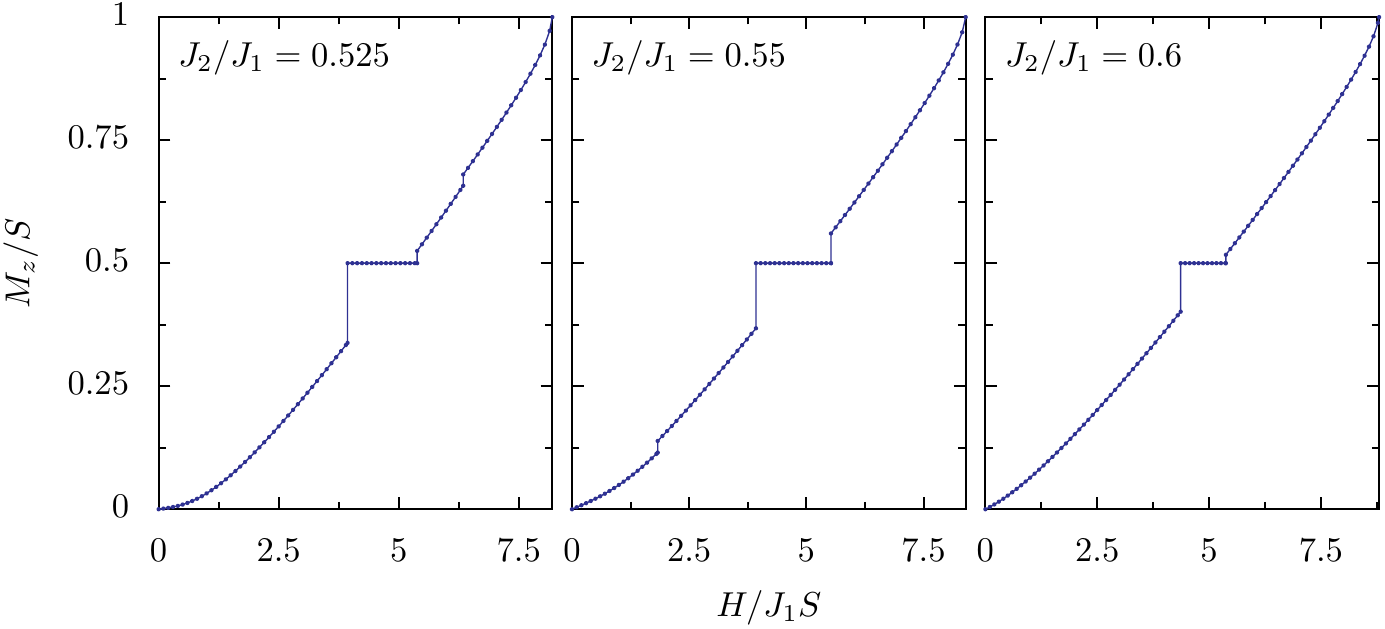}
\caption{Magnetization curves of the spin $1/2$ $J_1-J_2$ model for different ratios
$J_2/J_1$ obtained in the variational spin wave approach.}
\label{fig:Magnetization curves J1J2}
\end{figure}

The phase diagram obtained by comparing the ground-state energies for three relevant
spin structures is presented in Fig.~\ref{fig:J1J2 phase diagram} for $S=1/2$ and $S=1$.
The 1/2-magnetization plateau is stabilized by quantum fluctuations over a wide
range of parameters deep into the classically forbidden region $J_2/J_1>1/2$, though it
remains energetically unfavorable at $J_2/J_1=1/2$.
The width and position of the plateau are in good agreement with the exact diagonalization
results of finite clusters with up to $N=36$ sites \cite{Zhitomirsky-Honecker-Petrenko}.
Figure~\ref{fig:Magnetization curves J1J2} shows the magnetization curves for several ratios $J_2/J_1$.
The magnetization curve for $J_2/J_1=0.6$ with a large magnetization jump below the plateau and a much
smaller anomaly above the plateau is in good correspondence with the numerical data
for the same coupling ratio \cite{Zhitomirsky-Honecker-Petrenko}. For $J_2/J_1$ close to 1/2,
there is in addition a competition between the canted N\'eel and the canted stripe states. The
N\'eel state has a softer spectrum than the stripe state and is stabilized beyond its classical
boundary. This leads to an additional transition from the canted N\'eel state into the canted
stripe structure which shows up as a small jump either above ($J_2/J_1=0.525$) or below ($J_2/J_1=0.55$)
the 1/2-plateau.

It should be pointed out that other states than those considered may be stabilized.
One possible candidate, at the upper edge of the plateau, is the coplanar four sublattice state having
three classical spins parallel and the remaining spin pointing in a different direction. This state can be
naturally connected to the $\textit{uuud}$ state and is the analog of the state stabilized above the plateau
in the isotropic triangular lattice. However, such a structure could not be investigated in our linear spin-wave
approach since it is not a saddle point of the classical energy.

\begin{figure}[b]
 \centering
\includegraphics[width=\columnwidth]{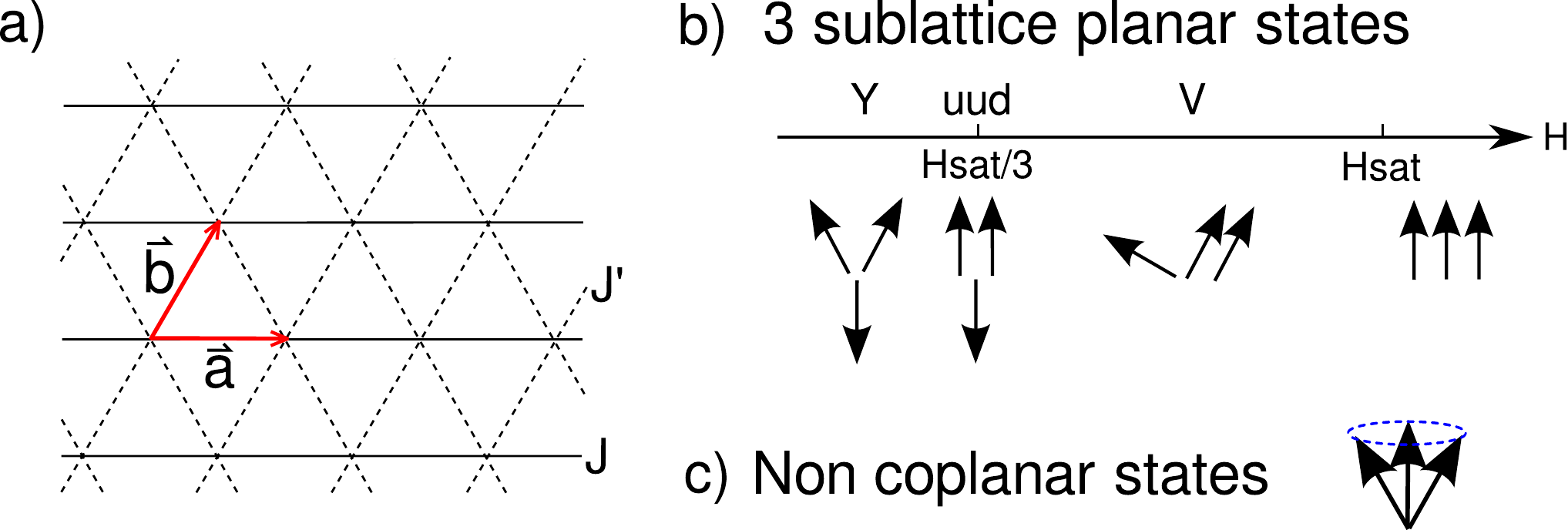}
\caption{a) Anisotropic triangular lattice and basis used in our calculations.
b) $3-$ sublattice planar structure as a function of the magnetic field.
c) Example of non coplanar canted helix.}
\label{fig:TriangularLattice}
\end{figure}

{\it Anisotropic triangular lattice.---}%
We now consider a second example of classically unstable magnetization plateau,
the nearest-neighbor Heisenberg antiferromagnet on an orthorhombically
distorted (anisotropic) triangular lattice. In this model spins are coupled by
$J_{ij}=J$ along horizontal chains and by $J_{ij}=J^\prime$ on zig-zag interchain bonds,
see Fig~\ref{fig:TriangularLattice}(a). The spin-1/2 model is relevant for
Cs$_2$CuBr$_4$ \cite{Ono2,Tsujii,Fortune09}, which has a $1/3$ magnetization plateau
although, with $J' \sim 0.75 J$, it is quite far form the isotropic limit.
The robustness of the $uud$ plateau in the $J$--$J'$  model has been studied
numerically \cite{Miyahara06} and analytically \cite{Alicea-Chubukov-Starykh}.
Nevertheless the extent of the plateau state around $H/H_{\textrm{sat}} \sim 1/3$ and $J'/J\sim 1$
and the nature of the states adjacent to the plateau region
are still open questions. Furthermore,  the stability method employed  by
Alicea {\it et al.}~\cite{Alicea-Chubukov-Starykh} allows to identify only
second-order transitions out of the plateau state, while experiments typically find
first-order transitions \cite{Fortune09}. We shall see below that this fact
finds a natural explanation in our theoretical approach.

The Fourier transform of the coupling interaction in the triangular-lattice, see
Fig.~\ref{fig:TriangularLattice}(a), is given by
$J_{\bf q}=2[J\cos{{\bf q} {\bf a}}+J^\prime\cos{{\bf q} {\bf b}}+J^\prime\cos{{\bf q}({\bf a}-{\bf b}})]$.
In zero field the classical ground state is a helical spin structure whose ordering wavevector
$\bf Q$ minimizes $J_{\bf q}$.
In the isotropic case $J=J^\prime$ this yields the well-known $120$\textdegree spin structure.
In the presence of a magnetic field the classical energy is minimized for canted helices or umbrella configurations,
see Fig.~\ref{fig:TriangularLattice}(c), which have
helical order in the $xy$ plane and uniform spin component in the field direction.
The canting angle of the helical structure measured with respect to the $z$ axis is given by
$\cos{\theta_{H}} = H/(J_0-J_{\bf Q})S$.
For the isotropic point $J^\prime/J=1$ the canted helical state is degenerate with the
coplanar Y- and $\mathbb{V}$-type structures, see Fig.~\ref{fig:TriangularLattice}(b).
Existing linear spin-wave calculations indicate that the coplanar structure is selected over
the non coplanar one in the isotropic lattice and that the \textit{uud} structure, classically stable at
the field $H_{\textrm{sat}}/3$, is stabilized by fluctuations over a finite field range \cite{Chubukov-Golosov}.

In the following we address the problem of the plateau stability for the anisotropic triangular
lattice model by comparing  ground-state energies for the canted helical state, the
\textit{uud} structure and the two $3$-sublattice planar states as a function of $J^\prime/J$ and
magnetic field.
Away from $J=J^\prime$, the $3$-sublattice planar structures turn out to be, 
like the \textit{uud}-state away from $H_{\textrm{sat}}/3$ and $J=J^\prime$, saddle points but not
local minima of 
the classical energy, and to compute their zero-point energies, we use
the variational spin-wave approach suggested above and add a staggered local field
to the Hamiltonian
\footnote{See the supplemental material (below) for details concerning the spin wave calculation
in the case of the anisotropic triangular lattice AFM}.
\begin{figure}[t]
 \centering
\includegraphics[width=0.8\columnwidth]{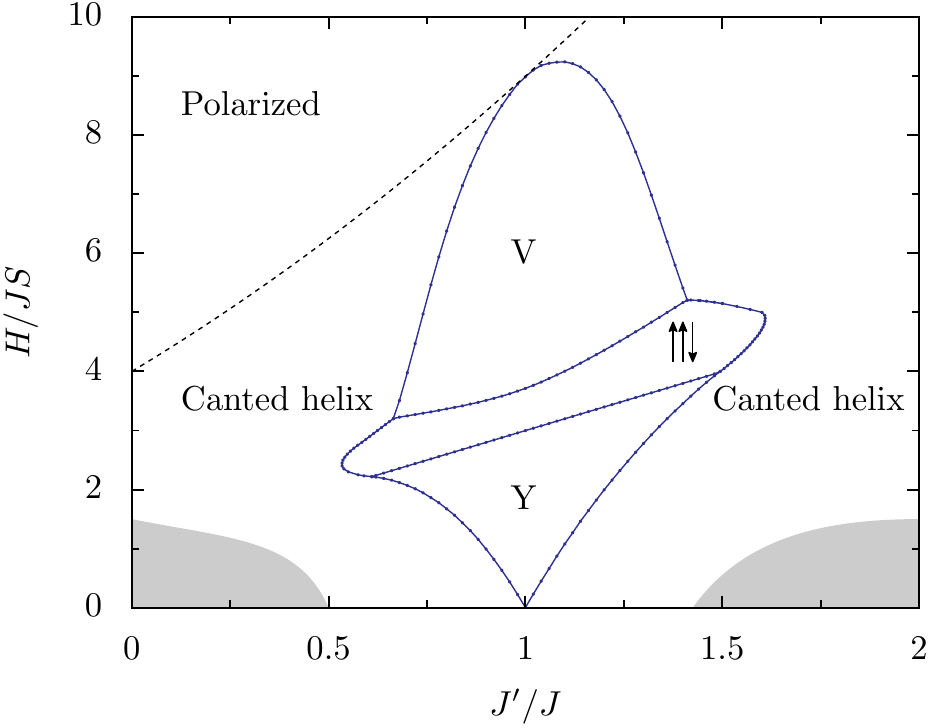}
\caption{Phase diagram of the spin-1/2 anisotropic triangular lattice in magnetic field. Y and V regions
denote  $3-$sublattice planar states. The dashed line is the classical saturation field. The gray shading
denotes regions where other phases than the canted helical states may be expected.}
\label{fig:Anisotropic Triangular lattice phase diagram}
\end{figure}

The resulting phase diagram is shown in
Fig.~\ref{fig:Anisotropic Triangular lattice phase diagram}.
The \textit{uud} plateau state is stabilized well beyond the isotropic limit and
extends over the range $0.5\lsim J^\prime/J \lsim 1.5$.
Coplanar states are stabilized above and below the magnetization plateau
with the exception of the plateau edges, where we find direct
first-order transitions from the $\textit{uud}$ state into the canted helical structure.
From the energetic comparison it appears that the \textit{uud} state does not extend into the Y-state region,
so that the corresponding portion of the lower boundary of the plateau is perfectly linear (see Fig.~\ref{fig:Anisotropic Triangular lattice phase diagram}).
This is almost certainly an artefact of the method, which only gives an upper bound to the
energy of the plateau, and the extent of the plateau is probably significantly larger. In fact, 
at $J=J^\prime$, we obtain a plateau width which is only half that predicted in Refs.~\cite{Chubukov-Golosov,Honecker04}. We also note that
the only coplanar states considered in our calculation are Y- and $\mathbb{V}$-type structures,
while for a substantial mismatch between $J$ and $J'$,
incommensurate coplanar structures may be also stabilized by quantum fluctuations.
The variational spin-wave approach is not well-suited for treating them and we only remark
that they may appear on the phase diagram at the expense of the canted helical structure.
Finally, the gray shading in Fig.~\ref{fig:Anisotropic Triangular lattice phase diagram} indicates regions,
where new quantum phases are expected.
In fact, in zero field, theoretical and numerical approaches point to collinear spin correlations
for weakly coupled chains, [\onlinecite{Starykh-Balents,White,Reuther-Thomale}],
while in the limit of strong interchain couplings the AFM N\'eel state should be stable down to
$J^\prime/J\approx 1.5$ \cite{Reuther-Thomale}.

As compared to those of Alicea {\it et al.}, who also predicted an extended plateau region for
small distortions [$(1-J'/J)^2\lesssim 0.3$] for the $S=1/2$ case~\cite{Alicea-Chubukov-Starykh},
our results bring in a number of new insights.
In the first place, the symmetry between $J'/J<1$ and $J'/J>1$ is lost.
Second, a transition out of the plateau into the canted helical states is clearly present.
Finally, for $J'/J=0.75$ relevant for Cs$_2$CuBr$_4$ \cite{Ono2,Tsujii,Fortune09},
we find a magnetization plateau width $\Delta H/H_{\rm sat} \approx 0.106$, significantly
larger than the experimental value $\Delta H/H_{\rm sat} \approx 0.052$ in Cs$_2$CuBr$_4$. Since
in most cases our approach underestimates the plateau width, the difference must be 
due to additional effects not included in the anisotropic model, for instance 
the competition between quantum effects and Dzyaloshinskii-Moriya interactions \cite{Griset11}.

{\it Conclusion.---}%
We have developed a general method to investigate the stabilization of classical magnetization plateaus
in cases where the corresponding configuration is not a minimum of the classical energy. This method
is extremely simple since it only relies on the diagonalization of quadratic bosonic Hamiltonians and does not
require to go beyond linear spin-wave theory, yet it appears to give remarkably accurate results, even for
spin $1/2$. This has been demonstrated in two cases of current
interest, the $J_1$--$J_2$ Heisenberg model on the square lattice and the Heisenberg model on the anisotropic triangular lattice, for which it predicts that plateaus at magnetization $1/2$ and $1/3$ respectively are stabilized over a wide range of parameters.

We acknowledge useful discussions with Andrey Chubukov, Sergey Korshunov and Karlo Penc.
This work has been supported by the Swiss National Fund and by MaNEP.

\bibliography{bibliography}


\section{Supplemental material: Linear spin wave theory for the anisotropic triangular lattice AFM in a magnetic field}

The general structure of the bosonic fluctuation Hamiltonian around a given classical state is given in Eq.~(\ref{eq:H sw}) of the main text.
In this supplemental material we present the specific expressions for the canted helical state, the \textit{uud} state and the
$3-$ sublattice coplanar structures.
\paragraph{Canted helical state ---}
The classical energy per site of the canted helical state expressed as a function of the canting angle $\theta$ is given by
\begin{equation}
E_{\textrm{cl}}^H=S^2(J_0\cos^2\theta+J_{\bf Q}\sin^2\theta)/2-HS\cos\theta.
\end{equation}
It is minimal for $\cos\theta^H=H/(J_0-J_{\bf Q})S$.
The term $\bf{\hat{a}}_{\bf k}^\dagger$ in Eq.~(\ref{eq:H sw}) of the main text denotes $(a_{\bf k}^\dagger,a_{-\bf k})$ and $M_{\bf k}$ is the $2\times2$ matrix 
\begin{equation}\label{eq:M helix}
M_{\bf k}({\bf Q})=\left(\begin{array}{cc}
			      A_{\bf k}({\bf Q})+C_{\bf k}({\bf Q}) & B_{\bf k}({\bf Q}) \\
			      B_{\bf k}({\bf Q}) & A_{\bf k}({\bf Q})-C_{\bf k}({\bf Q})
			      \end{array}\right)
\end{equation}
with coefficients
\begin{equation}\label{eq:Coeff M helix}
\begin{array}{l}
A_{\bf k}({\bf Q})=S\left(-J_{\bf Q}+\frac{1}{4}(\cos^2\theta+1)(J_{\bf k+Q}+J_{\bf k-Q})+\frac{1}{2}\sin^2\theta J_{\bf k}\right), \\[2mm]
B_{\bf k}({\bf Q})=S\left(\frac{1}{4}(\cos^2\theta-1)(J_{\bf k+Q}+J_{\bf k-Q})+\frac{1}{2}\sin^2\theta J_{\bf k}\right),\\[2mm]
C_{\bf k}({\bf Q})=S\cos\theta(J_{\bf k+Q}-J_{\bf k-Q})/2.
\end{array}
\end{equation}
The additional term $\Delta_{\bf k}$ in Eq.~(\ref{eq:H sw}) of the main text is given by $\Delta_{\bf k}=-SJ_{\bf Q}$.
\paragraph{$3-$sublattice coplanar states ---}
The classical energy per site of any 3- sublattice coplanar structure is given by
\begin{equation}\label{eq:E coplanar}
\begin{array}{lll}
 E_{\textrm{cl}}^\textrm{coplanar}&=&S^2(J+2J^\prime)(\cos{\alpha_{1,2}}+\cos{\alpha_{1,3}}+\cos{\alpha_{2,3}})/3 \\[2mm]
                            & &-SH(\cos{\alpha_1}+\cos{\alpha_2}+\cos{\alpha_3})/3
\end{array}
\end{equation}
where $\alpha_{i,j}=\alpha_i-\alpha_j$ are the spin orientations measured with respect to the field direction.
The angles $\alpha_i$ minimizing Eq.~(\ref{eq:E coplanar}) are given by
\begin{equation}
 \alpha_1^\textrm{Y}=\pi, \quad \cos\alpha_2^\textrm{Y}=\frac{1}{2}\left(\frac{H}{H_{\textrm{sat}}/3}+1\right),\quad \alpha_2^\textrm{Y}=-\alpha_3^\textrm{Y}
\end{equation}
for $0\leq H \leq H_{\textrm{sat}}/3$ and
\begin{equation}
\begin{array}{c}
\displaystyle \cos\alpha_1^\textrm{V}=\frac{H}{2H_\textrm{sat}}\left(3-\frac{H_\textrm{sat}^2}{H^2}\right), \\
\displaystyle \cos\alpha_2^\textrm{V}=\cos\alpha_3^\textrm{V}=\frac{H}{4H_\textrm{sat}}\left(3+\frac{H_\textrm{sat}^2}{H^2}\right)
\end{array}
\end{equation}
for fields in the range $H_{\textrm{sat}}/3 \leq H \leq H_{\textrm{sat}}$, where $H_{\textrm{sat}}$ is the saturation field $H_{\textrm{sat}}=3(J+2J^\prime)S$.

Since the states considered have three sites per unit cell the term $\bf{\hat{a}}_{\bf k}^\dagger$ in Eq.~(\ref{eq:H sw}) of the main text denotes 
$(a_{{\bf k},1}^\dagger,a_{{\bf k},2}^\dagger,a_{{\bf k},3}^\dagger,a_{{-\bf k},1},a_{{-\bf k},2},a_{{-\bf k},3})$ and $M_{\bf k}$ is the $6\times6$ matrix 
\begin{equation}\label{eq:M coplanar}
 M_{\bf k}^\textrm{Y,V}=\left(
\begin{array}{cccccc}
    A     & D_{\bf k}^\star & H_{\bf k}^\star      &     0     & E_{\bf k}^\star & I_{\bf k}^\star \\
D_{\bf k} &       B         & F_{\bf k}^\star      & E_{\bf k} &          0      & G_{\bf k}^\star \\
H_{\bf k} &    F_{\bf k}    &        C             & I_{\bf k} &      G_{\bf k}  & 0          \\
    0     & E_{\bf k}^\star & I_{\bf k}^\star      &     A     & D_{\bf k}^\star & H_{\bf k}^\star \\
E_{\bf k} &       0         & G_{\bf k}^\star      & D_{\bf k} &          B      & F_{\bf k}^\star \\
I_{\bf k} &    G_{\bf k}    &        0             & H_{\bf k} &      F_{\bf k}  & C
\end{array} 
\right)
\end{equation}
with
\begin{equation}\label{eq:Coeff M coplanar}
\begin{array}{l}
A=\left[-S(J+2J^\prime)(\cos{\alpha_{1,2}}+\cos{\alpha_{1,3}})+H\cos\alpha_1\right],\\[2mm]
B=\left[-S(J+2J^\prime)(\cos{\alpha_{1,2}}+\cos{\alpha_{2,3}})+H\cos\alpha_2\right],\\[2mm]
C=\left[-S(J+2J^\prime)(\cos{\alpha_{2,3}}+\cos{\alpha_{1,3}})+H\cos\alpha_3\right],\\[2mm]
D_{\bf k}=S(\cos\alpha_{1,2}+1)\gamma^{(1)}_{\bf k}/2,  \quad E_{\bf k}=S(\cos\alpha_{1,2}-1)\gamma^{(1)}_{\bf k}/2, \\[2mm]
F_{\bf k}=S(\cos\alpha_{2,3}+1)\gamma^{(1)}_{\bf k}/2,  \quad G_{\bf k}=S(\cos\alpha_{2,3}-1)\gamma^{(1)}_{\bf k}/2, \\[2mm]
H_{\bf k}=S(\cos\alpha_{1,3}+1)\gamma^{(2)}_{\bf k}/2,  \quad I_{\bf k}=S(\cos\alpha_{1,3}-1)\gamma^{(2)}_{\bf k}/2.
\end{array}
\end{equation}
where $\gamma^{(1)}_{\bf k}$ and $\gamma^{(2)}_{\bf k}$ are given by
\begin{equation}
 \begin{array}{l}
 \gamma^{(1)}_{\bf k}=Je^{i {\bf k}({\bf a}+{\bf b})}+J^\prime\left(1+ e^{i {\bf k}(-\bf{a}+2{\bf b}) }\right), \\[2mm]
 \gamma^{(2)}_{\bf k}=Je^{i {\bf k}(-{\bf a}+2{\bf b})}+J^\prime e^{i {\bf k}({\bf a}+{\bf b})}\left(1+ e^{i {\bf k}(-\bf{a}+2{\bf b}) }\right).
 \end{array}
\end{equation}
The additional term $\Delta_{\bf k}$ in Eq.~(\ref{eq:H sw}) of the main text is given by $\Delta_{\bf k}=A+B+C$.
\paragraph{\textit{uud} state ---}
The \textit{uud} state also belongs to the family of coplanar states.
Hence the expressions for the classical energy and for the coefficients of the fluctuation Hamiltonian 
(Eqs.~(\ref{eq:E coplanar}), (\ref{eq:M coplanar}) and (\ref{eq:Coeff M coplanar}))
obtained previously can be applied to the \textit{uud} state if one replaces $ \alpha_1=\pi$ and $ \alpha_2=\alpha_3=0$.

\end{document}